\documentclass[aps,showpacs,twocolumn,prd]{revtex4}

\usepackage{graphicx}
\usepackage{amsmath}

\newcommand{\intq}{\ensuremath{\int\frac{d^4 q}{(2\pi)^4}}}
\newcommand{\ppara}{\ensuremath{p_\parallel}}

\newcommand{\qpara}{\ensuremath{q_\parallel}}
\newcommand{\qperp}{\ensuremath{q_\perp}}

\newcommand{\gammapara}{\ensuremath{\gamma_\parallel}}
\newcommand{\nn}{\nonumber}

\begin{document}

\title{Dynamical electron mass in a strong magnetic field}

\author{Shang-Yung Wang}

\email[]{sywang@mail.tku.edu.tw}

\affiliation{Department of Physics, Tamkang University, Tamsui, Taipei 25137, Taiwan}

\date{December 20, 2007}

\begin{abstract}
Motivated by recent interest in understanding properties of strongly magnetized matter, we study the dynamical electron mass generated through approximate chiral symmetry breaking in QED in a strong magnetic field. We reliably calculate the dynamical electron mass by numerically solving the nonperturbative Schwinger-Dyson equations in a consistent truncation within the lowest Landau level approximation. It is shown that the generation of dynamical electron mass in a strong magnetic field is significantly enhanced by the perturbative electron mass that explicitly breaks chiral symmetry in the absence of a magnetic field.
\end{abstract}


\maketitle

Properties of matter in strong magnetic fields are of basic interest~\cite{Ferrer:2005vd,Mandal:2006fv,Shabad:2006gf,Campanelli:2005pn,Shabad:2007xu} and have great potential applications in the physics of compact stellar objects and the early universe cosmology~\cite{Grasso:2000wj}. The observations of soft gamma repeaters and anomalous X-ray pulsars have provided compelling evidence that the magnetic fields on the surface of young neutron stars are in the range of $10^{14}-10^{16}$~G~\cite{Thompson:1995gw}. It has been suggested that at the electroweak phase transition local magnetic fields as high as $10^{22}-10^{24}$~G could be generated~\cite{Brandenburg:1996fc}. Situations of even stronger magnetic fields may exist in extreme astrophysical and cosmological environments.

It has been established that the magnetic catalysis of chiral symmetry breaking is a nonperturbative universal phenomenon~\cite{Gusynin:1994re,Gusynin:1995gt,Lee:1997zj}. A strong magnetic field acts as a catalyst for chiral symmetry breaking, leading to the generation of a dynamical fermion mass even at the weakest attractive interaction between fermions. The hallmark of this effect is the dimensional reduction from $(3+1)$ to $(1+1)$ in the dynamics of fermion pairing in a strong magnetic field when the lowest Landau level (LLL) plays the dominant role. The realization of this phenomenon in the chiral limit in QED (i.e., QED with massless fermions) has been studied extensively in the literature over the past decade~\cite{Gusynin:1995gt,Lee:1997zj,Gusynin:1998zq}.

But until very recently, there has been no agreement on the correct calculation of the dynamical fermion mass generated through chiral symmetry breaking in QED in a strong magnetic field, and contradictory results have been found in the literature~\cite{Gusynin:1998zq,Gusynin:2002yi}. The resolution of the contradiction lies in the establishment of the gauge fixing independence of the dynamically generated fermion mass calculated in the nonperturbative Schwinger-Dyson (SD) equations approach~\cite{Leung:2005yq}. In particular, the study of Ref.~\cite{Leung:2005yq} has provided an unambiguous identification of the infinite subset of diagrams that contribute to chiral symmetry breaking in a strong magnetic field, and led to a consistent calculation of the dynamically generated fermion mass, reliable in the weak coupling regime and the strong field limit (for a brief review, see Ref.~\cite{Wang:2007bg}).

In order to highlight the most important physics regarding the mechanism of chiral symmetry breaking in a strong magnetic field, the phenomenon has been studied in the literature mostly in the chiral limit. Nevertheless, the universal nature of the phenomenon dictates that in realistic massive QED in a strong magnetic field, the electron will acquire a dynamical mass generated through the modification of the vacuum structure that is induced by the strong magnetic field. This effect is essentially analogous to that of the approximate chiral symmetry breaking in QCD and the Nambu--Jona-Lasinio model~\cite{Klevansky:1992qe}, where, in addition to the perturbative current quark masses, the quarks acquire nonperturbative constituent masses of dynamical origin that are brought about by the breaking of chiral symmetry.

In this article, we extend the study of Ref.~\cite{Leung:2005yq} to the case of massive QED in a strong magnetic field. Specifically, we consistently calculate the dynamically generated electron mass in the weak coupling regime and the strong field limit. While similar problems have been studied in the past~\cite{Gusynin:1998nh}, to the best of our knowledge, a consistent calculation of the dynamical electron mass in a strong magnetic field has not appeared in the literature. Apart from its theoretical interest, this problem is of practical interest and importance. In particular, sizable modifications of the electron mass as induced by strong magnetic fields will find applications in neutron star astrophysics and early universe cosmology.

The Lagrangian density of massive QED in an external magnetic field is given by
\begin{equation}
\mathcal{L}=-\frac{1}{4}F^{\mu\nu}F_{\mu\nu}
+\overline{\psi}\gamma^\mu[i\partial_\mu+e(A^\mathrm{ext}_\mu+A_\mu)]\psi
-m\overline{\psi}\psi,\label{L}
\end{equation}
where $\psi$ is the quantum fermion (electron) field, $A_\mu$ is the Abelian quantum gauge boson (photon) field, $F_{\mu\nu}$ is the corresponding electromagnetic field strength, and $A^\mathrm{ext}_\mu$ describes an external magnetic field. Here and henceforth, we set $\hbar=c=1$ and use the conventions in which $g_{\mu\nu}=\mathrm{diag}(-1,1,1,1)$ with $\mu,\nu=0,1,2,3$. The Dirac matrices satisfy $\{\gamma^\mu,\gamma^\nu\}=-2g^{\mu\nu}$ and
$\gamma^5=i\gamma^0\gamma^1\gamma^2\gamma^3$.

In the Lagrangian density \eqref{L}, we have not included the counterterms associated with the usual ultraviolet renormalization in QED. This is because we are solely interested in the dynamics of the electrons in the LLL, which is ultraviolet finite due to the effective dimensional reduction as remarked above. Hence, the constants $m$ and $e$ in the Lagrangian density \eqref{L} denote respectively the electron mass and the absolute value of its charge that are defined with an appropriate (perturbative) renormalization in the absence of external fields.

We choose the constant external magnetic field of strength $B$ in the $x_3$-direction. The corresponding vector potential is given by $A^\mathrm{ext}_\mu=(0,0,Bx_1,0)$ with $B>0$. A convenient formalism for the study of QED in the presence of a constant external magnetic field was developed a long time ago by Ritus~\cite{Ritus}. The so-called Ritus $E_p$ functions are constructed in terms of the simultaneous eigenfunctions (eigenvectors) of the mutually commuting operators $[\gamma^\mu(i\partial_\mu+eA^\mathrm{ext}_\mu)]^2$, $\Sigma^3\equiv i\gamma^1\gamma^2$ and $\gamma^5$, and form a complete set of Dirac matrix-valued orthonormal functions. The important advantage of the Ritus formalism is that in momentum space spanned by the $E_p$ functions, the Dirac equation for a noninteracting fermion in a constant external magnetic field is formally identical to that in the absence of external fields. It is noted that because the fermion mass $m$ is proportional to the identity operator, which obviously commutes with the above three operators, the Ritus formalism applies to both massless and massive QED.

It has been proved in Ref.~\cite{Leung:2005yq} that the bare vertex approximation (BVA) is a consistent truncation of the nonperturbative SD equations within the lowest Landau level approximation (LLLA). With a momentum independent fermion self-energy that fulfills the Ward-Takahashi (WT) identity in the BVA within the LLLA, it can be shown that (i) the truncated vacuum polarization is transverse; (ii) the truncated fermion self-energy is gauge independent when evaluated on the fermion mass shell. In particular, the would-be gauge dependent contribution to the truncated fermion self-energy, which arises from the gauge dependent term in the full photon propagator, vanishes identically on the fermion mass shell. As a consequence, the dynamical fermion mass, obtained as the solution of the truncated SD equations evaluated on the fermion mass shell, is manifestly gauge independent. The gauge independent analysis presented in Ref.~\cite{Leung:2005yq} is very general in nature and not specific to massless QED. Here we indicate the crucial points in the analysis extended to massive QED.

The motion of the LLL electrons is restricted in directions perpendicular to the magnetic field, leading to an effective dimensional reduction from $(3+1)$ to $(1+1)$ in the dynamics of fermion pairing in a strong magnetic field. Consistent with the WT identity in the BVA within the LLLA~\cite{Leung:2005yq}, the full propagator for the LLL electron in momentum space (spanned by the $E_p$ functions) is given by
\begin{equation}
G(\ppara)=\frac{1}{\gammapara\cdot\ppara+m_\ast}\,\Delta,\label{G}
\end{equation}
where $m_\ast$ is the (gauge independent) dynamical electron mass in a strong magnetic field, which should not be confused with the perturbative electron mass $m$ in the absence of a magnetic field. The dynamical electron mass $m_\ast$ is yet to be determined by solving the truncated SD equations self-consistently. In the above expression, $\ppara$ denotes the longitudinal momentum, namely, $\ppara^\mu=(p^0 ,p^3)$ and  $\Delta=(1+\Sigma^3)/2$ is the projection operator on the electron states with the spin polarized along the external magnetic field. The projection operator $\Delta$ satisfies the property $\Delta\,\gamma^\mu\,\Delta=\gammapara^\mu\,\Delta$,
which clearly reflects the effective dimensional reduction from $(3+1)$ to $(1+1)$ in the dynamics of the LLL electrons.

The WT identity in the BVA within the LLLA guarantees that the vacuum polarization $\Pi_{\mu\nu}(q)$ is transverse, i.e., $q_\mu\Pi^{\mu\nu}(q)=0$. An explicit calculation yields
\begin{equation}
\Pi^{\mu\nu}(q)=\Pi(\qpara^2,\qperp^2)
\biggl(g_\parallel^{\mu\nu}-\frac{q^\mu_\parallel
q^\nu_\parallel}{\qpara^2}\biggr),\label{Pimunu}
\end{equation}
where $\qpara^2=-q_0^2+q_3^2$ and $\qperp^2=q_1^2+q_2^2$. Eq.~\eqref{Pimunu} implies that the full photon propagator  in covariant gauges takes the form
\begin{eqnarray}
\mathcal{D}^{\mu\nu}(q)&=&\frac{1}{q^2+\Pi(\qpara^2,\qperp^2)}
\biggl(g_\parallel^{\mu\nu}-\frac{q^\mu_\parallel
q^\nu_\parallel}{\qpara^2}\biggr)+\frac{g_\perp^{\mu\nu}}{q^2}\nn\\
&&+\frac{q^\mu_\parallel q^\nu_\parallel}{q^2\qpara^2}
+(\xi-1)\frac{1}{q^2}\frac{q^\mu q^\nu}{q^2},\label{D}
\end{eqnarray}
where $\xi$ is the gauge fixing parameter with $\xi=1$ being the Feynman gauge. In the above expressions, the polarization function
$\Pi(\qpara^2,\qperp^2)$ is given by
\begin{gather}
\Pi(\qpara^2,\qperp^2)=\frac{2\alpha}{\pi}\,
eB\,\exp\biggl(-\frac{q_\perp^2}{2eB}\biggr)
F\biggl(\frac{\qpara^2}{4m_\ast^2}\biggr),\label{Pi}\\
F(u)=1-\frac{1}{2u\sqrt{1+1/u}}\,\log\frac{\sqrt{1+1/u}+1}{\sqrt{1+1/u}-1},
\end{gather}
where $\alpha=e^2/4\pi$ is the fine-structure constant. The dimensionless function $F(u)$ has the following asymptotic behavior: $F(u)\simeq 2u/3$ for $|u|\ll 1$, and $F(u)\simeq 1$ for $|u|\gg 1$. Hence, photons of momenta $m_\ast^2\ll|\qpara^2|\ll eB$ and $\qperp^2 \ll eB$ are screened with a characteristic screening length $\ell=1/\sqrt{(2\alpha/\pi)eB}$ induced by the strong magnetic field.

The self-energy of the LLL electron evaluated on the mass shell, $\ppara^2=-m_\ast^2$, leads to the so-called gap equation that determines the dynamical electron mass $m_\ast$ self-consistently. The WT identity in the BVA within the LLLA guarantees that contributions to the LLL electron self-energy from the gauge dependent term as well as from the terms proportional to $q^\mu_\parallel q^\nu_\parallel/\qpara^2$ in $\mathcal{D}^{\mu\nu}(q)$ vanish identically on the electron mass shell (see Ref.~\cite{Leung:2005yq} for a detailed discussion). Thus, in the BVA within the LLLA we obtain the (gauge independent) on-shell electron SD equation
\begin{eqnarray}
m_\ast&=&m+ie^2\intq\,\frac{m_\ast}{(p-q)_\parallel^2+m_\ast^2}\nn\\
&&\times\,\frac{\exp(-q_\perp^2/2eB)}{q^2+\Pi(\qpara^2,\qperp^2)} \bigg|_{\ppara^2=-m_\ast^2},\label{m}
\end{eqnarray}
where the projection operator $\Delta$ that multiplies both sides of the equation has been dropped. Using the mass shell condition $\ppara^\mu=(m_\ast,0)$ that corresponds to a LLL electron at rest and performing a Wick rotation to Euclidean space, we find the gap equation to be given by
\begin{eqnarray}
m_\ast&=&m+\frac{\alpha}{2\pi^2}\int
d^2\qpara\,\frac{m_\ast}{q_3^2+(q_4-m_\ast)^2+m_\ast^2}\nn\\
&&\times\int_0^\infty d\qperp^2\,
\frac{\exp(-q_\perp^2/2eB)}{\qpara^2+\qperp^2+
\Pi(\qpara^2,\qperp^2)}, \label{gap}
\end{eqnarray}
where $\qpara^2=q_3^2+q_4^2$ is the photon longitudinal momentum squared in Euclidean space.

Before proceeding further, we discuss the solution to the gap equation \eqref{gap} in the chiral limit (i.e., $m=0$). This will be useful for our discussions below. The solution in the chiral limit (denoted here and henceforth by $m_\mathrm{dyn}$) was obtained numerically and shown to be fitted by the analytic expression~\cite{Leung:2005yq}
\begin{equation}
m_\mathrm{dyn}=a\,\sqrt{2eB}\;\alpha\,\exp\left[-\frac{\pi}{\alpha\log(b/\alpha)}\right],
\label{mdyn}
\end{equation}
where $a$ is a constant of order one and $b\simeq 2.3$. From Eq.~\eqref{mdyn} it can be seen clearly that while on the one hand $m_\mathrm{dyn}$ scales as $\sqrt{2eB}$ and increases with increase of $B$, on the other hand it is exponentially suppressed at weak coupling. The wide separation of scales $m_\mathrm{dyn}\ll\sqrt{eB}$, together with the gauge independence of $m_\mathrm{dyn}$, is at the heart of the fact that the result of $m_\mathrm{dyn}$ given by Eq.~\eqref{mdyn} is reliable in the weak coupling regime and the strong field limit~\cite{Leung:2005yq}. To get a feeling of the order of magnitudes involved, it is instructive to note that for $\alpha=1/137$, $m_\mathrm{dyn}$ is about 34 orders of magnitude smaller than the energy between adjacent Landau levels, $\sqrt{2eB}$, and it would require an enormous magnetic field of about $10^{82}$ G to have $m_\mathrm{dyn}$ comparable to $m$. As a result, this nonperturbative effect can be safely ignored in the chiral limit in QED even though ultrastrong magnetic fields may be under consideration.

However, as will be seen below, in realistic massive QED the generation of dynamical electron mass in a strong magnetic field is significantly enhanced by the perturbative electron mass that explicitly breaks chiral symmetry in the absence of a magnetic field. This is the novel result of the present article.

\begin{figure}[t]
\begin{center}
\includegraphics[width=3.25in,keepaspectratio=true,clip=true]{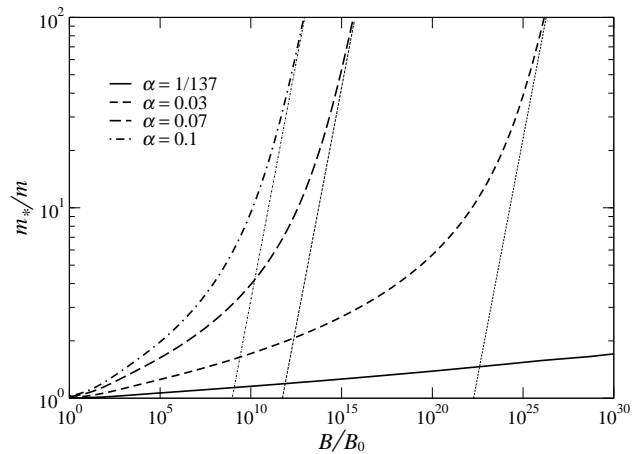}
\end{center}
\caption{Plot of the dynamical electron mass $m_\ast$ (in units of the perturbative electron mass $m$) as a function of the magnetic field strength $B$ (in units of the characteristic value $B_0=m^2/e\simeq 4.4\times 10^{13}$ G) for several values of the fine-structure constant $\alpha$. The thin lines represent the corresponding results in the chiral limit, i.e., $m_\mathrm{dyn}$. Note that for $\alpha=1/137$ the corresponding result in the chiral limit lies outside the plot range.}
\label{fig:mass}
\end{figure}

We have numerically solved the gap equation \eqref{gap} to obtain $m_\ast$ as a function of $B$. In Fig.~\ref{fig:mass} the dynamical electron mass $m_\ast$ (together with its value in the chiral limit, $m_\mathrm{dyn}$) is plotted against the magnetic field strength $B$ for several values of the fine-structure constant $\alpha$. While we are not able to find an analytic expression that fits the numerical results, an analysis of the gap equation \eqref{gap} shows that for fixed $\alpha$ its solution in an asymptotically strong magnetic field is reduced to the solution in the chiral limit. In other words, for fixed $\alpha$ we have $m_\ast\approx m_\mathrm{dyn}$ as $B\to\infty$. This asymptotic behavior is verified numerically as can be seen clearly in Fig.~\ref{fig:mass}.

Several important features of the dynamical electron mass generated in a strong magnetic field can be gleaned from Fig.~\ref{fig:mass}.
\begin{itemize}
\item[(i)]
There is a wide separation of scales $m_\ast\ll\sqrt{eB}$ as long as the coupling is weak and the magnetic field is strong. As remarked above, together with the gauge independence of $m_\ast$, the wide separation of scales means that our results for $m_\ast$ are reliable in the weak coupling regime and the strong field limit.

\item[(ii)]
Let $B_m$ denote the magnetic field for which $m_\mathrm{dyn}$ is equal to $m$ (i.e., the intercept of the thin line with the abscissa in Fig.~\ref{fig:mass}). It is clear from the figure that for $B\simeq B_m$, the corresponding $m_\ast$ is about one order of magnitude larger than $m$, a property that is fairly independent of the values of $\alpha$. This provides a distinct signature that the generation of dynamical electron mass in a strong magnetic field is significantly enhanced by the perturbative electron mass. Specifically, we note that for $\alpha=1/137$ there already has been a few percent increase in the electron mass around $10^{15}$ G, the typical magnetic fields on the surface of young neutron stars. Such an effect is within the precision of current and future astrophysical measurements. Furthermore, we note that the transition of the behavior of $m_\ast$ from the intermediate to the asymptotic takes place around $B\simeq B_m$, and again is fairly independent of $\alpha$.

\item[(iii)]
For fixed $B$, $m_\ast$ increases with increase of $\alpha$. At weak coupling, the dynamical contribution to the electron mass is small but sizable as compared to its exponentially suppressed counterpart in the chiral limit. Nevertheless, the dynamical contribution becomes substantial and eventually dominant over the perturbative electron mass as the coupling increases. This aspect is of particular importance when the effects of the running coupling in a strong magnetic field are taken into consideration.
\end{itemize}

The enhancement of the dynamical electron mass can be understood in terms of the screening effect modified by the perturbative electron mass, $m$. First, we consider the chiral limit. The corresponding polarization function $\Pi(\qpara^2,\qperp^2)$ is given by Eq.~\eqref{Pi} with the replacement $F(\qpara^2/4m_\ast^2)\to F(\qpara^2/4m_\mathrm{dyn}^2)$. In the region of $B$ where $m_\mathrm{dyn}$ is exponentially small, one can make the approximation $F(\qpara^2/4m_\mathrm{dyn}^2)\simeq F(\infty)=1$. When plugged into the gap equation \eqref{gap}, this in turn implies that photons with (Euclidean) momenta $0<\qpara^2\ll eB$ are effectively screened. The screening effect explains the smallness of $m_\mathrm{dyn}$. Away from the chiral limit, on the other hand, the perturbative electron mass $m$ introduces into the problem an additional energy scale that is independent of the LLL dynamics. In the region of $B$ where $\delta m_\ast\equiv m_\ast-m\ll m$ (as can be seen in Fig.~\ref{fig:mass}, this would be the same region of $B$ considered above in the chiral limit), one can make the replacement $F(\qpara^2/4m_\ast^2)\to F(\qpara^2/4m^2)$ in $\Pi(\qpara^2,\qperp^2)$. This in turn means that photons with momenta $m^2\ll\qpara^2\ll eB$ are effectively screened. As a result, the contribution from photons with momenta $0<\qpara^2\ll m^2$ that are not screened is responsible for the enhancement of the dynamical electron mass in a strong magnetic field.

The above arguments do not depend on the specific value of $\alpha$, and remain valid up to a magnetic field for which $\delta m_\ast/m\sim\mathcal{O}(1)$ [or, alternatively, $m_\mathrm{dyn}/m\sim\mathcal{O}(1)$]. As we have noticed above, this takes place around the magnetic field $B\simeq B_m$, for which $m_\ast$ is almost about one order of magnitude larger than $m$. This also explains why the transition of the behavior of $m_\ast$ from the intermediate to the asymptotic takes place around $B\simeq B_m$.

In conclusion, the significant enhancement of the dynamical electron mass in QED in a strong magnetic field is a novel effect. We envisage a similar enhancement of the dynamical quark masses in QCD in a strong magnetic field~\cite{Agasian:1999sx}. It would be interesting and useful to consider applications of these effects in astrophysics and cosmology.

I would like to thank C.\ N.\ Leung for a careful reading of the manuscript. This work was supported in part by the National Science Council of Taiwan under grant 96-2112-M-032-005-MY3.

\end{document}